\documentclass[twocolumn,showpacs,preprintnumbers,amsmath,amssymb]{revtex4}
\usepackage{dcolumn}
\usepackage{bm}
\usepackage{amsmath,amssymb,amsfonts,latexsym,fancyhdr,graphicx}

\newcommand{\ket}[1]{\displaystyle{|#1\rangle}}
\newcommand{\bra}[1]{\displaystyle{\langle#1|}}
\newcommand{\al}{\alpha}
\newcommand{\Om}{\Omega}
\newcommand{\om}{\omega}

\newcommand{\G}{\Gamma}
\newcommand{\g}{\gamma}
\newcommand{\si}{\sigma}
\newcommand{\la}{\lambda}

\begin{document}
\title{Exact dynamics of entanglement and entropy in structured environments}
\author{L. Mazzola, S. Maniscalco, J. Piilo, and K.\,-\,A. Suominen}
\affiliation{Department of Physics and Astronomy, University of
Turku, FI-20014 Turun yliopisto, Finland}

\begin{abstract}
We study the exact entanglement dynamics of two qubits interacting
with a common zero-temperature non-Markovian reservoir. We consider
the two qubits initially prepared in Bell-like states or extended
Werner-like states. We study the dependence of the entanglement
dynamics on both the degree of purity and the amount of entanglement
of the initial state. We also explore the relation between the
entanglement and the von Neumann entropy dynamics and find that
these two quantities are correlated for initial Bell-like states.
\end{abstract}
\maketitle

\section{Introduction}

Entanglement is a key resource in modern quantum information theory
and technology. Entangled states play a central role in quantum key
distribution, superdense coding, quantum teleportation and quantum
error correction \cite{Nielsen}. However, realistic quantum systems
are never completely isolated from their surroundings. The
inevitable interaction between a system and its environment leads to
decoherence phenomena and degradation of entanglement \cite{Breuer}.

Once entanglement has been lost, it cannot be restored by local
operations \cite{Plenio}. It is therefore important to understand
the process of disentanglement in order to control the effects of
noise and to preserve entanglement. For these reasons the study of
entanglement dynamics of quantum systems in realistic situations has
become of increasing importance.

Recently much attention has been devoted to the process of
finite-time disentanglement, also known as \lq\lq entanglement
sudden death\rq\rq (ESD), in a bipartite system \cite{Yu}. This
phenomenon consists in the complete disappearance of the bipartite
entanglement in a finite time, as opposed to the time-evolution of
the coherences of the single parts which vanish only asymptotically.

ESD sets a limit on the life-time and usability of entanglement for
practical purposes. Hence, lots of efforts have been done in order
to understand the conditions under which ESD
occurs~\cite{YuScience}. Finite-time disentanglement has been found
in different physical systems, such as qubits or harmonic
oscillators~\cite{Paz}. ESD appears when the two parts of the system
interact with either independent environments~\cite{Yu,Agarwal} or a
common one~\cite{Ficek}. The first studies within the Markov
approximation have been extended to non-Markovian environments,
where the memory of the reservoir adds revivals to entanglement
dynamics~\cite{Bellomo,NoiESD}.

In Ref.~\cite{NoiESD} we have studied entanglement sudden death when
two-qubits are prepared in a Bell-like state with two excitations,
and entanglement sudden birth for qubits prepared in a separable
state. Here we focus on the dynamics of a class of states having an
\lq\lq X\rq\rq-structure density matrix, namely the extended
Werner-like states (EWL). This class of states plays a crucial role
in many applications of quantum information theory, such as
teleportation~\cite{AdhikariHorodecki} and quantum key distribution
\cite{Acin}. Moreover, such a choice will give us the chance to
study how the entanglement dynamics and its revivals are related to
the purity and the amount of entanglement of the initial state. We
also focus on the interplay between entanglement and mixedness for
the qubits states. We investigate the connection between these two
quantities, comparing concurrence and von Neumann entropy dynamics
for initial Bell-like states, and finding clear correlations between
them.

The paper is organized as follows. In Sec. II we review the exactly
solvable model of two qubits interacting with a common Lorentzian
structured reservoir. In Sec. III we study the entanglement and von
Neumann entropy dynamics for initial Bell-like states and we prove
that sudden death of entanglement can never occur if the qubits are
initially in a mixed state having at most one excitation. In Sec. IV
we focus on the time evolution of EWL states, and compare our
results with those obtained in Ref.~\cite{BellomoEWL} for
independent structured reservoirs. Finally, we summarize our results
in Sec. V.

\section{The model}
In this section we describe the model we use to study the dynamics
of two two-level systems (qubits) interacting with a common
zero-temperature bosonic reservoir. Our approach is non-Markovian
and non-perturbative, i.e., it does not rely on either the Born or
the Markov approximations~\cite{NoiESD}.

The Hamiltonian of the system, in the rotating wave approximation,
is given by $H=H_{0}+H_{\mathrm{int}}$,
\begin{equation}
   H_{0}=\om_{0}(\si_{+}^{A}\si_{-}^{A}+\si_{+}^{B}\si_{-}^{B})
   +\sum_{k}\om_{k}a_{k}^{\dag}a_{k},\label{H0bare}
\end{equation}
\begin{equation}\label{Hintbare}
   H_{\mathrm{int}}=(\si_{+}^{A}+\si_{+}^{B})\sum_{k}g_{k}a_{k}+\mathrm{h.c. },
\end{equation}
where $\si_{\pm}^{A}$ and $\si_{\pm}^{B}$ are the Pauli raising and
lowering operators for qubit A and B respectively, $\om_{0}$ is the
Bohr frequency of the two identical qubits, $a_{k}$ and
$a_{k}^{\dagger}$, $\om_{k}$ and $g_{k}$ are the annihilation and
creation operators, the frequency and the coupling constant of the
field mode $k$, respectively.

In order to solve the dynamics of the two qubits we need to specify
the properties of the environment. In the following we assume that
the two qubits interact resonantly with a non-Markovian Lorentzian
structured reservoir, such as the electromagnetic field inside a
lossy cavity \cite{Harbook}, having spectral distribution

\begin{equation}
   J(\omega)=\frac{\Omega^2}{2\pi}\frac{\Gamma}{(\om-\om_{0})^2+(\Gamma/2)^2},
\end{equation}
where $\Gamma$ is the width of the Lorentzian function and $\Omega$
the coupling strength.

We are mainly interested in the dynamics of entanglement and in the
effects that the non-Markovian reservoir induces on the correlation
between the two qubits. To quantify entanglement we use the Wootters
concurrence~\cite{Wootters}, defined as
$C(t)=\textrm{max}\{0,\sqrt{\la_{1}}-\sqrt{\la_{2}}-\sqrt{\la_{3}}-\sqrt{\la_{4}}\}$,
where $\{\la_{i}\}$ are the eigenvalues of the matrix
$R=\rho(\si_{y}^{A}
\otimes\si_{y}^{B})\rho^{\ast}(\si_{y}^{A}\otimes\si_{y}^{B})$, with
$\rho^{\ast}$ denoting the complex conjugate of $\rho$ and
$\si_{y}^{A/B}$ are the Pauli matrices for atoms $A$ and $B$. This
quantity attains its maximum value of 1 for maximally entangled
states and vanishes for separable states.

We focus now on the dynamics of initial \lq\lq X\rq\rq\ states. We
use the method described in Ref.~\cite{NoiESD} to calculate the time
evolution of the density matrix
\begin{equation}\label{rhot}
   \rho(t)=\left(
      \begin{array}{cccc}
        a(t) & 0 & 0 & w(t) \\
        0 & b(t) & z(t) & 0 \\
        0 & z^{*}(t) & c(t) & 0 \\
        w^{*}(t) & 0 & 0 & d(t) \\
      \end{array}
   \right),
\end{equation}
which is written in the basis
$\{\ket{00},\ket{10},\ket{01},\ket{11}\}$. Due to the structure of
the differential equations for the density matrix elements [see
Eqs.~(5) and (6) in Ref.~\cite{NoiESD}], the \lq\lq X\rq\rq\ form is
preserved during the evolution. In the Appendix we present the
analytical solution in the Laplace transform space for a particular
type of \lq\lq X\rq\rq\ state.

For this class of states the concurrence assumes a simple analytic
expression
\begin{equation}\label{conc}
   C(t)=\mathrm{max}\{0,C_{1}(t),C_{2}(t)\},
\end{equation}
where
\begin{equation}\begin{split}\label{C1C2}
   C_{1}(t)&=2|w(t)|-2\sqrt{b(t) c(t)},\\
   C_{2}(t)&=2|z(t)|-2\sqrt{a(t) d(t)}.
\end{split}\end{equation}
We notice that coherences give a positive contribution to $C_{1}(t)$
and $C_{2}(t)$ and so to concurrence, while the negative parts
involve populations only.

In the next section we will also consider the evolution of the
mixedness of the two qubit state, which we quantify through the von
Neumann entropy, defined as
\begin{equation}\label{VonNeumann}
S(\rho)=-\mathrm{Tr}\{\rho(t)\ln(\rho(t))\}.
\end{equation}
Von Neumann entropy is equal to zero for pure states, and attains
its maximum value (equal to $\ln N$ with $N$ the dimension of the
Hilbert space) for a maximally mixed state.

\section{Entanglement and Von Neumann entropy dynamics for
Bell-like states}

Here, we seek for the connection between the dynamics of
entanglement and von Neumann entropy of two qubits prepared in
Bell-like states
\begin{equation}\label{Phi}
   \ket{\Phi}=\al\ket{10}+e^{i\theta}(1-\al^2)^{1/2}\ket{01},
\end{equation}
and
\begin{equation}\label{Psi}
   \ket{\Psi}=\al\ket{00}+e^{i\theta}(1-\al^2)^{1/2}\ket{11}.
\end{equation}

Our aim is to understand the interplay between these two different
physical quantities, in particular when peculiar phenomena such as
ESD or \lq\lq entanglement sudden birth\rq\rq
(ESB)~\cite{FicekESB,Solano,NoiESD}, and revivals of entanglement
occur.

The entanglement dynamics of two qubits in a Lorentzian structured
reservoir, prepared in a Bell-like state with two excitations as in
Eq.~\eqref{Psi}, has been presented in Ref.~\cite{NoiESD}. There we
have studied in detail the difference with the common Markovian
reservoir case and the independent reservoirs non-Markovian case.
The evolution of entanglement for a Bell-like state as in
Eq.~\eqref{Phi} has been studied also in Ref.~\cite{Zeno}.

For both the Bell-like states in Eqs.~\eqref{Phi} and \eqref{Psi}
the entanglement dynamics is the result of two combined effects: the
backaction of the non-Markovian reservoir and the reservoir-mediated
interaction between the qubits.

The memory effects due to the non-Markovianity of the reservoir
causes oscillations in entanglement dynamics. These oscillations are
typical also of the independent reservoirs case~\cite{Bellomo}, but
they disappear completely for Markovian reservoirs~\cite{Ficek}. The
sharing of the reservoir plays also a special role. Indeed, the
common reservoir provides an effective coupling between the qubits,
and so consequently creates quantum correlations between them. As a
result, qubits prepared in a factorized state can become entangled
due to the interaction with the common reservoir. This is in
contrast with the independent reservoirs case in which a factorized
state of the qubits can never evolve into an entangled state.

The results in Ref.~\cite{Zeno} show that ESD does not occur for a
Bell-like state with one excitation as in Eq.~\eqref{Phi} for any
value of $\al^2$. ESD does not appear for the same Bell-like state
even if the dipolar interaction between the qubits is included
\cite{Dipolar}. Actually, a straightforward calculation shows that
for every pure or mixed state of the qubits containing at most one
excitation, ESD and ESB cannot take place. In fact, the density
matrix describing a generic mixed state with maximum one excitation,
written in the same basis of Eq.~\eqref{rhot}, has the form
\begin{equation}\label{density1}
   \rho(t)=\left(
      \begin{array}{cccc}
        a(t) & j(t) & k(t) & 0 \\
        j^{*}(t) & b(t) & z(t) & 0 \\
        k^{*}(t) & z^{*}(t) & c(t) & 0 \\
        0 & 0 & 0 & 0 \\
      \end{array}
   \right).
\end{equation}
The expression for the concurrence for any value of the parameters
is
\begin{equation}
C(t)=\mathrm{max}\{0,2|z(t)|\}.
\end{equation}
Here concurrence is directly given by the coherence between the
$\ket{10}$ and $\ket{01}$ states. Since the coherence vanishes in
asymptotic way, there cannot be ESD for any generic state with
maximum one excitation. Analogously, entanglement can be smoothly
generated but it cannot suddenly appear. This result does not depend
on the degree of purity of the state. This is true as long as the
form of the density matrix in Eq.~\eqref{density1} is maintained. On
the other hand, if some population is transferred to the two
excitations state then ESD can appear. This is the case of two
qubits in a Bell state interacting with a non-RWA common
reservoir~\cite{Jung}.

One problem we want to address here is to understand how the amount
of entanglement changes with the purity of the state. In the next
section we will explore this aspect by preparing the qubits in a
particular class of mixed states. Furthermore, it is useful to see
how the mixedness of the initial state changes with time because of
the interaction with the environment, and if there exists some
connection between the von Neumann entropy and entanglement
dynamics.

In order to provide a clear interpretation of the dynamics, it is
useful to recall the four-state effective description~\cite{NoiESD}.
Indeed, the state of the two qubits is equivalent to a four-state
system, in which three states $\{\ket{00},\
\ket{+}=(\ket{10}+\ket{01})/\sqrt{2},\ \ket{11}\}$ interact in
ladder configuration with the electromagnetic field, and the fourth
state $\{\ket{-}=(\ket{10}-\ket{01})/\sqrt{2}\}$ is completely
decoupled from the other states and the electromagnetic field. The
states $\ket{+}$ and $\ket{-}$ are known as super-radiant and
sub-radiant states, respectively.

\begin{figure}[!]
\begin{center}
\includegraphics[width=8.6cm]{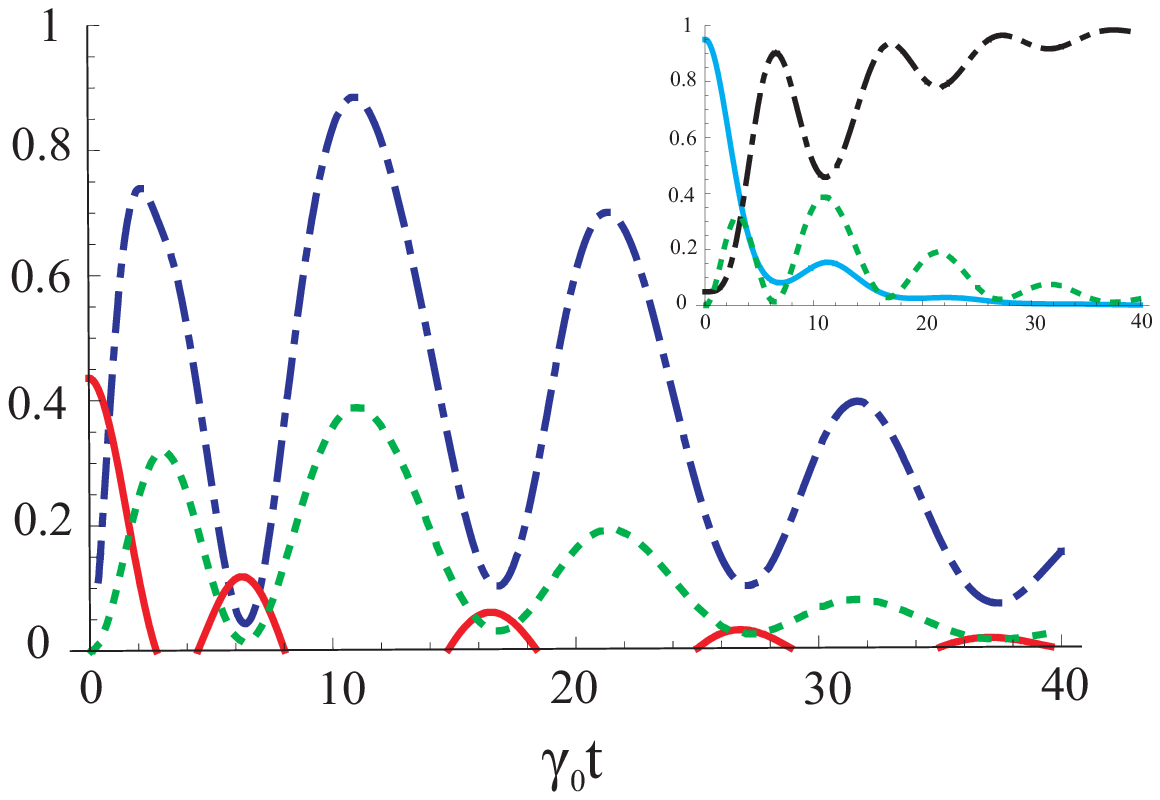}
\end{center}
\caption{(Color online) Dynamics in a common Lorentzian structured
reservoir as a function of scaled time for two atoms prepared in the
Bell-like state $\Psi$ with $\al^{2}=1/20$ and $\theta=0$. Solid red
line is concurrence; dotted-dashed blue line is von Neumann entropy;
dotted green line is the population of the super-radiant state
$\rho_{++}(t)$. In the inset: dotted green line is the super-radiant
state population; dotted-dashed black line is the ground state
population; solid light blue line is the excited state $\ket{11}$
population.}\label{fig:VNCBell2}
\end{figure}

\begin{figure}[!]
\begin{center}
\includegraphics[width=8.6cm]{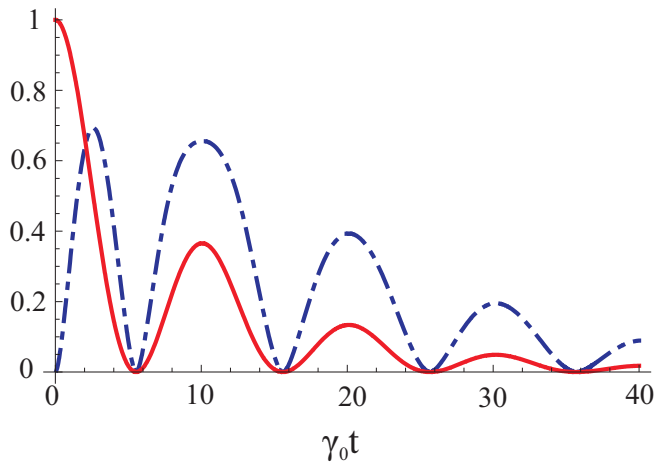}
\end{center}
\caption{(Color online)  Dynamics in a common Lorentzian structured
reservoir as a function of scaled time for two atoms prepared in the
Bell-like state $\Phi$ with $\al^{2}=1/2$ and $\theta=0$. Solid red
line is concurrence and dotted-dashed blue line is von Neumann
entropy, whiich overlaps with the super-radiant state
population.}\label{fig:VNCBell1}
\end{figure}

\begin{figure}[!]
\begin{center}
\includegraphics[width=8.6cm]{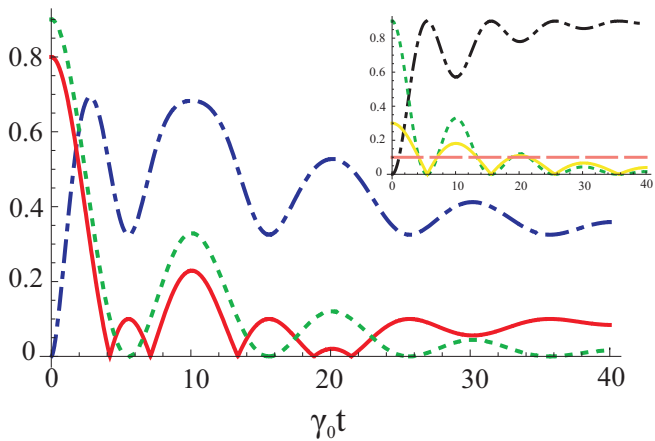}
\end{center}
\caption{(Color online) Dynamics in a common Lorentzian structured
reservoir as a function of scaled time for two atoms prepared in the
Bell-like state $\Phi$ with $\al^{2}=1/5$ and $\theta=0$. Solid red
line is concurrence; dotted-dashed blue line is von Neumann entropy;
dotted green line is super-radiant state population. In the inset:
dotted green line is super-radiant state population; dotted-dashed
black line is ground state population; dashed pink line is
sub-radiant state population; solid yellow line is absolute value of
the coherence between super-radiant and sub-radiant
states.}\label{fig:VNCBell12}
\end{figure}

\begin{figure}[!]
\begin{center}
\includegraphics[width=8.6cm]{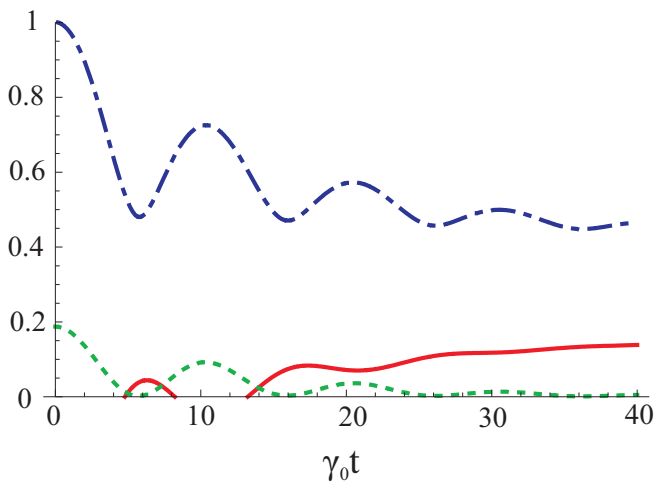}
\end{center}
\caption{(Color online) Dynamics in a common Lorentzian structured
reservoir as a function of scaled time for two atoms prepared in the
mixed state \eqref{ESBinistate} with $\al^{2}=0.75$. Solid red line
is concurrence; dotted-dashed blue line is von Neumann entropy;
dotted green line is super-radiant state population.
}\label{fig:VNCESB}
\end{figure}

We look at the evolution of the von Neumann entropy of the qubit
pair when the state is initially prepared in a Bell-like state of
the form \eqref{Phi} and \eqref{Psi}. It is particularly interesting
to see how the degree of purity of the state evolves when the qubits
undergo ESD. Since ESD occurs when $\alpha^2\lesssim1/4$
\cite{NoiESD}, we choose $\alpha^2=1/20$ and $\theta=0$ for the sake
of convenience. In Fig.~\ref{fig:VNCBell2} we notice that ESD
appears for high value of mixedness of the system (entropy peaks
coincide with minima of concurrence). Moreover, the revivals of
entanglement appear roughly in correspondence of the minimum of the
von Neumann entropy, when the state becomes purer. A careful
analysis shows that the dynamics of entropy follows the population
of the super-radiant state $\rho_{++}(t)$, (in the Appendix we
provide the analytic expression of the von Neumann entropy for this
initial state). As Fig.~\ref{fig:VNCBell2} shows, they attain their
relative maxima and minima at the same times. Moreover, we can
deduce from Eqs.~\eqref{conc} and \eqref{C1C2} that entanglement
dynamics is a function of the population of the super-radiant state,
which is $\rho_{++}(t)=2\sqrt{b(t)c(t)}$ for the particular initial
state of Eq.~\eqref{Psi}. Therefore whenever the population
$\rho_{++}(t)$ reaches its relative maxima, the state attains a
maximum value of mixedness, the time-dependent part of the
concurrence ($2|w(t)|-\rho_{++}(t)$) becomes negative and
entanglement disappears. On the other hand, whenever the population
of the super-radiant state reaches a minimum, the population of the
$\ket{11}$ excited state and the $\ket{00}$ ground state have their
maxima (see inset in Fig.~\ref{fig:VNCBell2}), and the system goes
toward a Bell-like state $\ket{\Psi}$. As a consequence the system
becomes purer and entanglement is partially recovered.

When the system is prepared in a Bell-like state as the one of
Eq.~\eqref{Phi} there is no ESD, however it is still of interest to
understand how the revivals of entanglement are related to the
degree of purity of the state. For the Bell state $\ket{+}$
entanglement has exactly the same dynamics of the population of the
super-radiant state, as shown in Fig.~\ref{fig:VNCBell1}. Moreover
the zeroes of entanglement and entropy coincide. For those times, in
fact, the system goes into the ground state which is pure and
factorized. When some population returns in the super-radiant state
entanglement is recovered, and the state is again mixed.

When $\alpha^2\neq1/2$ and/or $\theta\neq0$ the initial state is a
superposition of super-radiant and sub-radiant states. Although the
sub-radiant state does not evolve in time, being decoupled from the
super-radiant and ground states, it affects the entanglement and
entropy dynamics. For the sake of convenience we choose $\al^2=1/5$
and $\theta=0$. In Fig.~\ref{fig:VNCBell12} we see that entropy
still follows closely the time evolution of the super-radiant state
population, having its relative minima in the same positions of the
zeroes of $\rho_{++}(t)$, (the analytic expression of the von
Neumann entropy is in the Appendix). On the contrary, entanglement
has new relative maxima when the population of the super-radiant
state is zero. This is due to the presence of the sub-radiant state.
In fact, in this case both the super-radiant and the sub-radiant
states contribute to the total entanglement. Thus there are two
different sets of entanglement maxima, those associated with the
maxima of the super-radiant state population, and those associated
to the sub-radiant state. Entanglement is zero whenever the
population of the super-radiant state $\rho_{++}(t)$, the population
of the sub-radiant state $\rho_{--}(t)$, and the absolute value of
the coherence between super-radiant and sub-radiant states
$\rho_{+-}(t)$, are equal. This can be explained in the light of the
expressions in Eqs.~\eqref{eqAppendix} in the Appendix, where
$b(t)$, $c(t)$ and $z(t)$ are written as a function of those
quantities. In fact, when $\rho_{++}(t)$, $\rho_{--}(t)$ and
$|\rho_{+-}(t)|$ have equal value, then $z(t)$ is equal to zero,
irrespective of the sign of $\rho_{+-}(t)$. Specifically, this
happens in correspondence to the 1st, 4th and 5th zeroes of
entanglement, where $b(t)=0$ and $c(t)=2\rho_{--}=k$. In this case
the state of the system becomes
$(1-k)\ket{00}\bra{00}+k\ket{01}\bra{01}$, which is clearly not
entangled. Similarly, for the 2nd and 3rd zeroes, $b(t)=k$, $c(t)=0$
and the non-entangled state is
$(1-k)\ket{00}\bra{00}+k\ket{10}\bra{10}$.

To conclude this section we consider the following mixed and
factorized initial state
\begin{equation}\begin{split}\label{ESBinistate}
   \rho(0)&=(\al^{2}\ket{0_{A}}\bra{0_{A}}+(1-\al^{2})\ket{1_{A}}\bra{1_{A}})\\
   &\quad\otimes (\al^{2}\ket{0_{B}}\bra{0_{B}}+(1-\al^{2})\ket{1_{B}}\bra{1_{B}}).
   \end{split}
\end{equation}
The dynamics of this state in a common non-Markovian reservoir is
characterized by ESB and revivals of disentanglement, as we have
shown in Ref.~\cite{NoiESD}. In Fig.~\ref{fig:VNCESB} we see how
these interesting features are related to the degree of purity of
the state. As we have seen before, the positions of maxima and
minima of the entropy and of the super-radiant state population
match. The sudden creation of entanglement happens roughly when the
entropy hits its first minimum. Entanglement is again lost when the
amount of mixedness increases, and it reappears again when the
entropy reaches another minimum.

\section{Werner state}

In this section we study the two qubits  entanglement dynamics of
extended Werner-like states in a common zero-temperature Lorentzian
structured reservoir. Our goal here is to study how, starting from
an initial state that is not perfectly pure, as in realistic
experimental conditions, affects the entanglement dynamics, and in
particular the occurrence of ESD and ESB phenomena.

\subsection{Terminology and previous works}
The standard two-qubit Werner state, introduced in 1989 by Werner
\cite{Werner}, is defined as
\begin{equation}\label{Werner}
\rho_{W}=r\ket{-}\bra{-}+\frac{1-r}{4}\mathbb{I},
\end{equation}
where $\ket{-}$ is the singlet state. In Ref.~\cite{Werner} Werner
demonstrated that while pure entangled states always violate the
Bell inequality, mixed entangled states might not. The Werner state
is the first entangled state to be proven not violating any Bell
inequalities \cite{Werner}. The generalized or Werner-like states
are defined as
\begin{equation}\label{WernerL}
\rho_{WL}=r\ket{M}\bra{M}+\frac{1-r}{4}\mathbb{I}
\end{equation}
with $\ket{M}$ one of the four maximally entangled Bell states. For
a given $r$, Werner and Werner-like states exhibit the same
entanglement. A further generalization is the extended Werner-like
states (EWL), containing a non-maximally entangled state part, which
are defined as
\begin{equation}\label{EWL1}
\rho_{EWL}^{\Phi}=r\ket{\Phi}\bra{\Phi}+\frac{1-r}{4}\mathbb{I},
\end{equation}
where $\ket{\Phi}=\al\ket{10}+e^{i\theta}(1-\al^2)^{1/2}\ket{01}$,
and
\begin{equation}\label{EWL2}
\rho_{EWL}^{\Psi}=r\ket{\Psi}\bra{\Psi}+\frac{1-r}{4}\mathbb{I},
\end{equation}
with $\ket{\Psi}=\al\ket{00}+e^{i\theta}(1-\al^2)^{1/2}\ket{11}$.

The dynamics of entanglement of two qubits prepared in Werner,
Werner-like or extended Werner-like states has attracted a lot of
attention. The appearance of ESD has been demonstrated for two
qubits prepared in a Werner-like state and interacting with
Markovian independent reservoirs \cite{YuEberly,YuWerner} or with
independent noisy channels \cite{CJShan}. Violation of Bell
inequality has been examined in independent thermal reservoirs
\cite{Miranowicz,Wildfeuer}. Entanglement has been studied in a
Markovian common thermal reservoir \cite{ShangBinLi1}, in the
presence of collective dephasing \cite{ShangBinLi2} and in the
maximal noise limit \cite{Jamroxmax}. Finally, the dynamics of an
EWL in two independent Lorentzian structured reservoirs has been
investigated in Ref.~\cite{BellomoEWL}.

Moreover, Werner and Werner-like states have been used so far in
many applications in quantum information processes such as
teleportation \cite{AdhikariHorodecki} and entanglement
teleportation \cite{Lee}. The experimental preparation and
characterization of the Werner states have also been widely
investigated. Werner states are prepared via spontaneous parametric
down-conversion \cite{Zhang} or using a universal source of
entanglement \cite{Barbieri}, and used in ancilla-assisted process
tomography \cite{Altepeter} and secure quantum key distribution
\cite{Acin}.

\subsection{Entanglement dynamics}
We evaluate the dynamics of the entanglement of EWL states as a
function of the initial amount of mixedness, controlled by the
purity parameter $r$, and as a function of the initial degree of
entanglement measured by $\al^{2}$.

Looking at the evolution of this kind of states gives the
possibility to study how the degree of purity of the initial state
influences the entanglement dynamics. We also compare our results
with those obtained in Ref.~\cite{BellomoEWL} for two qubits in two
independent Lorentzian structured reservoirs.

\begin{figure}[!]
\begin{center}
\includegraphics[width=8.6cm]{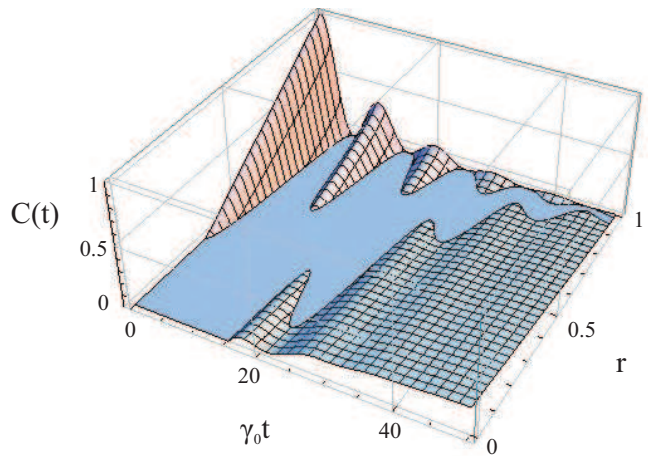}
\end{center}
\caption{(a) Concurrence as a function of scaled time and $r$ for
two atoms prepared in the Werner state~\eqref{EWL1} ($\Phi$) with
$\alpha^{2}=1/2$ and $\theta=0$.}\label{fig:CW1r1}
\end{figure}

\begin{figure}[!]
\begin{center}
\includegraphics[width=8.6cm]{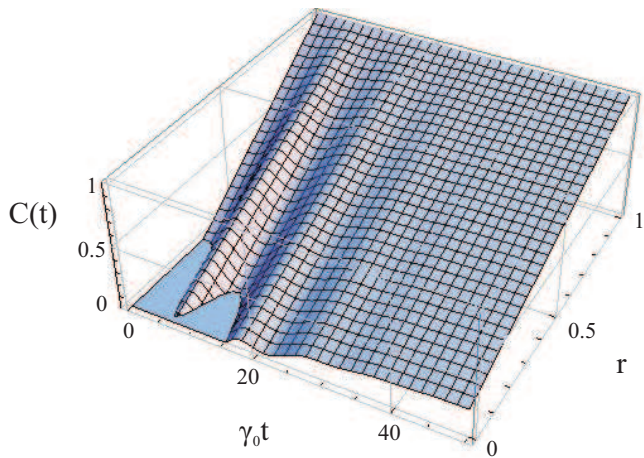}
\end{center}
\caption{(a) Concurrence as a function of scaled time and
$\alpha^{2}$ for two atoms prepared in the Werner state~\eqref{EWL1}
($\Phi$) with $r=1/2$ and $\theta=\pi$.}\label{fig:CW1Pir}
\end{figure}

\begin{figure}[!]
\begin{center}
\includegraphics[width=8.6cm]{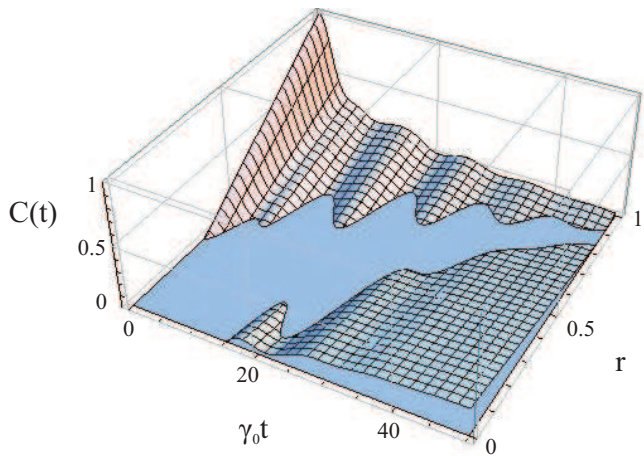}
\end{center}
\caption{(a) Concurrence as a function of scaled time and $r$ for
two atoms prepared in the Werner state~\eqref{EWL2} ($\Psi$) with
$\alpha^{2}=1/2$ and $\theta=0$.}\label{fig:CW2r}
\end{figure}

Figures~\ref{fig:CW1r1}, \ref{fig:CW1Pir} and \ref{fig:CW2r} show
the entanglement dynamics as a function of the dimensionless
quantity $\gamma_{0}t$ (with $\g_{0}=4\Om^{2}/\Gamma$ the Markovian
decay rate of the atoms) and of the purity parameter $r$, for the
two EWL states in Eqs.~\eqref{EWL1} and \eqref{EWL2}. In
Fig.~\ref{fig:CW1r1} the qubits are initially in the state given by
Eq.~\eqref{EWL1} with $\al^2=1/2$ and $\theta=0$. When $r=1$ the
qubits are prepared in the super-radiant state, entanglement
exhibits oscillations and ESD is never present. However, whenever a
little amount of mixdness is added, periods of finite-time
disentanglement immediately occur. This is due to the appearance of
some population in the excited state $\ket{11}$. As a consequence
the oscillating part of the concurrence
$\{2|z(t)|-2\sqrt{a(t)d(t)}\}$ can become negative. Due to the
non-Markovianity of the system and to the effective coupling
provided by the common reservoir, ESD regions are followed by
revivals, and for long times entanglement reaches a stationary
value. When $r<1/3$ the state is initially factorized, but as time
passes, because of the reservoir-mediated interaction, entanglement
between the qubits is suddenly created. The non-Markovianity of the
reservoir enriches the dynamics causing eventually revivals of
disentanglement. Eventually the entanglement reaches a non-zero
stationary value. The amount of entanglement that has been created
depends on the population of the sub-radiant state, which carries
the entanglement.

If the qubits are prepared in the state given by Eq.~\eqref{EWL1}
with $\al^2=1/2$ and $\theta=\pi$, the dynamics is completely
different compared to the previous $\theta=0$ case, as shown in
Fig.~\ref{fig:CW1Pir}. In fact when $r=1$ the qubits are prepared in
the sub-radiant state and concurrence does not evolve, being at any
time equal to 1. When the mixed part is present little oscillations
appear, but eventually the concurrence attains the stationary value
$(1+3r)/4$. When $r<1/3$ the initial state is factorized,
entanglement is suddenly created, a revival of disentanglement is
present, and again after some oscillations concurrence reaches its
stationary value.

When $\al^2\neq1/2$ and/or $\theta\neq0,\pi$ the non-mixed part of
the initial state is a superposition of super-radiant and
sub-radiant states. Thus the entanglement dynamics depends on the
weights of those two states. As a result there is a wide variety of
entanglement dynamics in between the two asymptotic behaviors
described above.

In Fig.~\ref{fig:CW2r} the qubits are prepared in the state given by
Eq.~\eqref{EWL2} with $\al^2=1/2$ and $\theta=0$. Note however that
anyway the results are independent of the choice of the relative
phase $\theta$. For $r=1$ the initial state is the Bell state
$(\ket{00}+\ket{11})/\sqrt{2}$. For this initial condition not only
there is no ESD, but also concurrence vanishes only for infinite
time. When an increasing amount of mixedness is present in the
initial state, finite-time disentanglement appear. ESD is then
followed by revivals and, as expected, a certain amount of
entanglement is preserved. Analogously to the other EWL state, for
$r<1/3$ the initial state is factorized. For the same reasons
previously mentioned, entanglement is suddenly created, momentarily
deteriorated, and it finally goes to the stationary value $r/4$,
coinciding with the population of the sub-radiant state. For this
type of EWL state different choices of $\al$ lead to the same
qualitative behavior of the entanglement dynamics.

The results we just described for two qubits in a common structured
reservoir are quite different from those presented in
Ref.~\cite{BellomoEWL} for two qubits in two independent Lorentzian
structured reservoirs. First of all, if two uncoupled qubits
interact with two independent reservoirs entanglement cannot be
created from a factorized state. Hence, the ESB region,
characterizing the dynamics in a common reservoir, is not present in
the results in Ref.~\cite{BellomoEWL}, for any of the EWL states.
Moreover, for qubits prepared in the initial state of
Eq.~\eqref{EWL2}, we notice that the reservoir-mediated interaction
between the qubits keeps the value of concurrence higher compared to
the two independent reservoirs case.

The crucial difference between the common and independent reservoirs
cases is that for qubits in two independent reservoirs the
decoherence-free sub-radiant state does not exist. When the qubits
are in two independent reservoirs, the entanglement dynamics of the
$\ket{+}$ and $\ket{-}$ states are the same. As a consequence the
relative phase $\theta$ in Eqs.~\eqref{Phi} and \eqref{EWL1} does
not affect the results. This is definitely in contrast with the
results we present in Figs.~\ref{fig:CW1r1} and \ref{fig:CW1Pir},
showing two completely different asymptotic behavior when changing
the relative phase $\theta$ of a $\pi$ factor.

\section{Conclusive Remarks}
In this paper we have investigated the connection between
entanglement and entropy dynamics in a system of two qubits
interacting with a common zero temperature non-Markovian reservoir.
We have used the exactly solvable model presented in
Ref.~\cite{NoiESD}, in the Appendix we attach the analytical
solution in the Laplace transform space in the case of EWL states.

We have compared the entanglement and von-Neumann entropy
time-evolution for two qubits prepared in a Bell-like state. We have
noticed that ESD seems to appear when the state becomes highly
mixed, whereas revivals of entanglement are associated to minima of
the entropy, where the state becomes purer. On the other hand, when
starting from a factorized state, sudden birth of entanglement
occurs for lower values of mixedness, while revivals of
disentanglement are accompanied by peaks of the entropy.

For a Bell-like state with one excitation the picture seems to be
more complex. However, by studying the dynamics of the population of
the super-radiant and sub-radiant states, we have realized that two
sets of maxima of entanglement can be identified. The super-radiant
state mainly carries the entanglement when it is maximally
populated; when its population is zero, the entanglement is
associated to the sub-radiant state. When the populations of the
states are equal, the entanglement vanishes.

We have also considered the entanglement dynamics of a particular
class of mixed states, the extended Werner-like states. We have
demonstrated that the amount of purity of the initial state plays a
key role in the entanglement dynamics, controlling the appearance of
ESD and ESB phenomena. For stronger non-Markovian conditions the
dynamics exhibits stronger and longer lasting entanglement
oscillations and an increasing number of dark periods and revivals
as well. On the other hand, in the Markovian regime, no oscillations
are present, however the basic features of the dynamics, and so the
ESD and ESB regions, are still present.

In general, entanglement and mixedness are two different physical
quantities characterizing the degree of non-classicality of a
quantum state. We think that it is important also from fundamental
point of view to understand the interplay between entanglement and
entropy, and that the present results shed new light on their
dynamical relation.

\acknowledgements

We thank Barry Garraway for stimulating discussions. This work has
been financially supported by M. Ehrnrooth Foundation,
V\"{a}is\"{a}l\"{a} Foundation, Turku University Foundation, Turun
Collegium of Science and Medicine, and the Academy of Finland
(projects 108699, 115682, 115982).

\appendix*
\section{}

Here we present the exact analytic solution for two qubits
interacting with a Lorentzian structured reservoir, when the qubits
are prepared in an EWL state as in Eqs.~\eqref{EWL1} and
\eqref{EWL2}. We provide the expressions for the density matrix
element in Eq. \eqref{rhot} as a function of the solution
$\tilde{\rho}_{ij}$ of the pseudomode master equation \cite{Barry97}
in Eqs.~(5) and (6) of Ref.~\cite{NoiESD}
\begin{equation}\begin{split}\label{eqAppendix}
a(t)&=\tilde{\rho}_{aa}(t)+\tilde{\rho}_{bb}(t)+\tilde{\rho}_{cc}(t),\\
b(t)&=\frac{\tilde{\rho}_{++}(t)+\tilde{\rho}_{--}(t)+\tilde{\rho}_{+-}(t)+\tilde{\rho}_{-+}(t)}{2},\\
c(t)&=\frac{\tilde{\rho}_{++}(t)+\tilde{\rho}_{--}(t)-\tilde{\rho}_{+-}(t)-\tilde{\rho}_{-+}(t)}{2},\\
z(t)&=\frac{\tilde{\rho}_{++}(t)-\tilde{\rho}_{--}(t)-\tilde{\rho}_{+-}(t)+\tilde{\rho}_{-+}(t)}{2},\\
d(t)&=1-a(t)-b(t)-c(t)=\tilde{\rho}_{ff}(t),\\
w(t)&=\tilde{\rho}_{af}(t),\\
\tilde{\rho}_{++}(t)&=\tilde{\rho}_{dd}(t)+\tilde{\rho}_{ee}(t).
\end{split}\end{equation}
The subscripts $a$, $b$ and $c$ refer to the states in which both
the qubits are in the ground state, and the pseudomode has 0, 1 or 2
excitations respectively. In $d$ and $e$ the atomic system is in the
super-radiant state, and the pseudomode has 0 or 1 excitations. $f$
is the state having 0 excitations in the pseudomode and both the
qubits in their excited states.

The analytic expressions are given in the Laplace transform space.
In the following we provide the solutions for a EWL state of the
form \eqref{EWL1}:
\begin{widetext}
\begin{equation}\begin{split}
\tilde{\rho}_{aa}(t)=\mathcal{L}^{-1}\Bigl\{-\frac{1}{4s}\bigl(-1+r+
\frac{64\G^{2}\Om^{4}(-1+r)l(s)}{j(s)k(s)}- \frac{8\G\Om^2(1+r+4
\al\sqrt{1-\al^2} r \cos{\theta})}{j(s)}\bigl)\Bigl\},
\end{split}\end{equation}
\begin{equation}\begin{split}
\tilde{\rho}_{bb}(t)=\mathcal{L}^{-1}\Bigl\{\frac{2
\Om^{2}}{j(s)}\bigl(1+r-\frac{8\G\Om^{2}(-1+r)l(s)}{k(s)}+4
\al\sqrt{1-\al^2}r\cos{\theta}\bigl)\Bigl\},
\end{split}\end{equation}
\begin{equation}
\tilde{\rho}_{cc}(t)=\mathcal{L}^{-1}\Bigl\{-\frac{48\Om^4(-1+r)(\G+s)}{k(s)}\Bigl\},
\end{equation}
\begin{equation}\begin{split}
&\tilde{\rho}_{dd}(t)=\mathcal{L}^{-1}\Bigl\{\frac{1}{4j(s)}\bigl(-\frac{1}{k(s)}(8\G\Om^{2}(-1+r)
(6\G^{5}+31\G^{4}s+20\G^{3}(-2\Om^{2}+3s^{2})+5\G^{2}s(-20\Om^{2}+11s^{2})\\&+8\G
(56\Om^{4}-\frac{17}{2}\Om^{2}s^{2}+3s^{4})+4(120\Om^4s-2\Om^2s^3 +
s^5)))+(8\Om^2+(\G+s)(\G+2s))(1+r+4\al\sqrt{1-\al^2}r\cos{\theta})\bigl)\Bigl\},
\end{split}\end{equation}
\begin{equation}
\tilde{\rho}_{ee}(t)=\mathcal{L}^{-1}\Bigl\{-\frac{2\Om^2(-1+r)(6\G^3+8\G\Om^2+
13\G^2 s+12\Om^2 s+9\G s^2+2 s^3)}{k(s)}\Bigl\},
\end{equation}
\begin{equation}\begin{split}
\tilde{\rho}_{ff}(t)=&\mathcal{L}^{-1}\Bigl\{-\frac{1}{4k(s)}\bigl((-1+r)(6\G^5+31\G^4
s+4\G^3 (38\Om^2+15s^2)+\G^2
s(412\Om^2+55s^2)+\\&8\G(40\Om^4+\frac{91}{2}\Om^2
s^2+3s^4)+4(72\Om^4 s+26\Om^2 s^3+s^5))\bigl)\Bigl\},
\end{split}\end{equation}
\begin{equation}
\tilde{\rho}_{+-}(t)=\mathcal{L}^{-1}\Bigl\{\frac{r(\G+2s)(-1+2\al^2+2
i\al\sqrt{1-\al^2}\sin{\theta}))}{2(4\Om^2+s(\G+2s)}\Bigl\},
\end{equation}
\end{widetext}
\begin{equation}
\tilde{\rho}_{-+}(t)=\tilde{\rho}^{*}_{+-}(t),
\end{equation}
\begin{equation}
\tilde{\rho}_{--}(t)=\frac{r(1-2\al\sqrt{1-\al^2}\cos{\theta})}{2}+\frac{1-r}{4},
\end{equation}
\begin{equation}
\tilde{\rho}_{af}(t)=0.
\end{equation}

The solutions for both the EWL states are written as functions of
$k(s)$, $j(s)$, $l(s)$ defined as
\begin{equation}\begin{split}
k(s)=(16\G\Om^{2}+2\G^{2}s+24\Om^{2}s+3\G
s^{2}+s^{3})\\(3\G^{3}+28\G\Om^{2}+11\G^{2}s+24\Om^{2}s+12\G
s^{2}+4s^{3}),
\end{split}\end{equation}
\begin{equation}
j(s)=(\G+2s)(8\Om^2+s(\G+s)),
\end{equation}
\begin{equation}
l(s)=(6\G^3+31\G^2 s+\G(56\Om^{2}+45s^2)+20(3\Om^2s+s^3)).
\end{equation}

Here we present the solutions for a EWL state of the form
\eqref{EWL2}:
\begin{widetext}
\begin{equation}\begin{split}
\tilde{\rho}_{aa}(t)=\mathcal{L}^{-1}\Bigl\{\frac{-1}{4s}\bigl(-1+r-4\al^2
r+\frac{8\G\Om^2(-1+r)}{j(s)}+ \frac{64\G^{2}\Om^{4}(-1+(-3+4
alp)r)l(s)}{j(s)k(s)}\bigl)\Bigl\},
\end{split}\end{equation}
\begin{equation}\begin{split}
\tilde{\rho}_{bb}(t)=\mathcal{L}^{-1}\Bigl\{\frac{2
\Om^{2}}{j(s)}\bigl(1-r-\frac{8\G\Om^{2}(-1+(-3+4\al^2)r)l(s)}{k(s)}\bigl)\Bigl\},
\end{split}\end{equation}
\begin{equation}
\tilde{\rho}_{cc}(t)=\mathcal{L}^{-1}\Bigl\{-\frac{48\Om^4(-1+(-3+4
\al^2)r)(\G+s)}{k(s)}\Bigl\},
\end{equation}
\begin{equation}\begin{split}
\tilde{\rho}_{dd}(t)=\mathcal{L}^{-1}\biggl\{-\frac{1}{4j(s)}\Bigl((8\Om^2+(\G+s)(\G+2s))(-1+r)+\frac{1}{k(s)}\bigl(8\G\Om^{2}(-1+(-3+4
\al^2)r)(6\G^{5}+31\G^{4}s\\+20\G^{3}(-2\Om^{2}+3s^{2})+5\G^{2}s(-20\Om^{2}+11s^{2})+8\G
(56\Om^{4}-\frac{17}{2}\Om^{2}s^{2}+3s^{4})+4(120\Om^4 s-2\Om^2 s^3
+ s^5))\bigl)\Bigl)\biggl\},
\end{split}\end{equation}
\begin{equation}
\tilde{\rho}_{ee}(t)=\mathcal{L}^{-1}\Bigl\{-\frac{2\Om^2(-1+(-3+4
\al^2)r)(6\G^3+8\G\Om^2+ 13\G^2 s+12\Om^2 s+9\G s^2+2
s^3)}{k(s)}\Bigl\},
\end{equation}
\begin{equation}\begin{split}
\tilde{\rho}_{ff}(t)=&\mathcal{L}^{-1}\Bigl\{-\frac{1}{4k(s)}\bigl((-1+(-3+4
\al^2)r)(6\G^5+31\G^4 s+4\G^3 (38\Om^2+15s^2)+\G^2
s(412\Om^2+55s^2)+\\&8\G(40\Om^4+\frac{91}{2}\Om^2
s^2+3s^4)+4(72\Om^4 s+26\Om^2 s^3+s^5))\bigl)\Bigl\},
\end{split}\end{equation}
\begin{equation}
\tilde{\rho}_{af}(t)=\mathcal{L}^{-1}\Bigl\{\frac{\al
r\sqrt{1-\al^2}((8\Om^2+(\G+s)(\G+2s))}{4\G\Om^2+\G^2 s+12\Om^2+3\G
s^2+2 s^3}\Bigl\},
\end{equation}
\end{widetext}
\begin{equation}
\tilde{\rho}_{--}(t)=\tilde{\rho}_{+-}(t)=\tilde{\rho}_{-+}(t)=0.
\end{equation}

We also provide the expressions of the von Neumann entropy for the
Bell-like states in Eqs.~\eqref{Phi} and \eqref{Psi}
\begin{widetext}
\begin{equation}\begin{split}
S_{\Psi}=&\frac{1}{2}\Bigl((a(t)+d(t))Log[4]
-2\tilde{\rho}_{++}(t)Log[\rho_{++}(t)]-(a(t)+d(t)-f(t))(Log[a(t)+d(t)-f(t)])\\&-(a(t)+d(t)+f(t))Log[a(t)+d(t)+f(t)]\Bigl),
\end{split}\end{equation}
\begin{equation}\begin{split}
S_{\Phi}=&\frac{1}{2}\Bigl(-2
a(t)Log[a(t)]-(\tilde{\rho}_{++}(t)+\tilde{\rho}_{--}(t)-j(t))Log[\tilde{\rho}_{++}(t)+\tilde{\rho}_{--}(t)-j(t)]\\
&-(\tilde{\rho}_{++}(t)+\tilde{\rho}_{--}(t)+j(t))Log[\tilde{\rho}_{++}(t)+\tilde{\rho}_{--}(t)+j(t)]\Bigl).
\end{split}\end{equation}
\end{widetext}

\end{document}